\documentclass[letterpaper,english,amssymb]{revtex4}

\usepackage{graphicx}
\usepackage{amssymb}
\usepackage{overpic}

\usepackage{babel}

\newcommand{\beq}{\begin{equation}}
\newcommand{\eeq}{\end{equation}}
\newcommand{\bea}{\begin{eqnarray}}
\newcommand{\eea}{\end{eqnarray}}

\newcommand{\kb}{{\bf k}}
\newcommand{\qb}{{\bf q}}

\begin{document}

\title{Renormalization-group approach to superconductivity: from weak to
  strong electron-phonon coupling}
\author{S.-W.~Tsai$^1$, A.~H.~Castro~Neto$^1$, R.~Shankar$^2$, and D.~K.~Campbell$^1$}

\affiliation
{$^1$ Department of Physics, Boston University, Boston, MA 02215\\
$^2$ Sloane Laboratory of Physics, Yale University, New Haven, CN 06520}

\date{\today}

\begin{abstract}
We present the numerical solution of the renormalization group (RG) equations
derived in Ref.\onlinecite{Tsai}, for the problem of superconductivity in the 
presence of both electron-electron and electron-phonon coupling at zero
temperature. We  study the instability of a Fermi liquid to a superconductor 
and the RG flow of the couplings in presence of retardation effects and
the crossover from weak to strong coupling. We show that our numerical
results provide an ansatz for the analytic solution of the problem in the asymptotic
limits of weak and strong coupling. 
\end{abstract}

\maketitle

The renormalization-group approach to interacting fermions in more than one
spatial dimension \cite{shankar} has been extensively applied to the study of
instabilities of the Fermi liquid state, and has become a major tool in the
study of correlated electron systems. Nevertheless, even for weak
electron-electron interactions, the picture is far from being complete, since
electrons in solids also interact with bosonic modes such as
phonons. Therefore, the development of an RG scheme \cite{Tsai} that includes
both electron-electron and electron-phonon interactions on an equal footing is 
an important advance. Experimental evidence show that in many strongly
correlated systems, such as high-temperature superconductors and organic
charge transfer salts, both electron interactions and phonons seem to play
an important role\cite{shen,organics,mgb2,millis,c60}.

The renormalization-group approach to interacting fermions coupled to phonons
was presented in Ref. [\onlinecite{Tsai}].  This approach takes retardation
effects and the 
presence of multiple energy scales fully into account. For a circular Fermi
surface, 
the RG equations predict the onset of the superconducting instability  in
agreement 
withe Eliashberg's superconducting theory \cite{eliashberg}. A large-$N$
analysis, 
where $N \approx E_F/\Lambda$ is the number of patches in the Fermi surface
($E_F$ 
is the Fermi energy and $\Lambda$ is the size of the patch), 
shows that in this case Eliashberg theory is asymptotically exact and Migdal's
theorem \cite{migdal} emerges as  a consequence of the $1/N$ expansion. 
Here we present a numerical solution 
of the RG equations at $T=0$, showing how the couplings, which are functions of
frequencies, flow with the RG procedure. In the limits of weak and strong
phonon coupling, simple analytical expressions are extracted on the basis
of the numerical solution. 

For completeness, in Sec. \ref{sec:derivation} the RG equations for the
self-energy and 
interaction couplings are derived. In Sec. \ref{sec:flows} we present the 
numerical results, and the analytical expressions associated with the
asymptotic limits. 
Sec. \ref{sec:conclusion} contains the concluding remarks.

\section{Derivation of the RG equations} 
\label{sec:derivation}

The action that describes electron-electron and electron-phonon interactions
can be  
written as $S(\psi,\phi) = S_e(\psi) + S_{ph}(\phi) +
S_{e-ph}(\psi,\phi) + S_{e-e}(\psi)$, 
where $\phi$ are bosonic fields, $\psi$ are fermionic (Grassman) fields (we
use units such that $\hbar = 1 = k_B$),  
\begin{eqnarray}
S_e = \int_{\omega \kb \sigma}
\psi^{\dagger}_{k \sigma} (i\omega - \epsilon_{\kb}) \psi_{k \sigma} \, ,
\end{eqnarray}
is the free electron action, $\epsilon_{\kb}$ is the electron dispersion as
a function of momentum $\kb$, and
\begin{eqnarray}
S_{ph} = \int_{\Omega \qb} \phi^{\dagger}_{q} (i\Omega - w_{\qb})
\phi_{q} \, ,
\end{eqnarray}
is the free phonon action where $w_{\qb}$ is the phonon dispersion
($\sigma=\uparrow,\downarrow$ is the electron spin, 
$k = \{\omega, \kb\}$ and $q = \{\Omega, \qb\}$,
where $\omega,\Omega$ are fermionic and bosonic Matsubara frequencies,
respectively, and $\kb,\qb$ are the momenta). 
The electron-phonon interaction can be written as:  
\begin{eqnarray}
S_{e-ph} =
\int_{\omega \kb \sigma} \int_{\Omega \qb} g(q) \psi^{\dagger}_{k+q \sigma}
\psi_{k \sigma}(\phi_q + \phi^{\dagger}_{-q}) \, , 
\end{eqnarray}
where $g(q)$ is the electron-phonon coupling constant. The electron-electron
interactions have the form:
\begin{eqnarray}
S_{e-e} = \frac{1}{2}
\prod_{i=1}^3 \int_{\omega_i \kb_i \sigma \sigma'} u(k_4,k_3,k_2,k_1)
\psi^{\dagger}_{k_4 \sigma} \psi_{k_2 \sigma} \psi^{\dagger}_{k_3
  \sigma^{\prime}} \psi_{k_1 \sigma^{\prime}}  \, , 
\end{eqnarray} 
where $u(\{k_i\})$ is a general spin-independent electron-electron coupling
which depends on the momenta and frequencies of the electrons, and
$k_4=k_1+k_2-k_3$ due to momentum conservation. 

Since the bosonic action is quadratic in the boson fields,  they can be
integrated  
out exactly, leading to an effective electron-electron problem with retarded
interactions:
\begin{eqnarray}
\tilde{u}(k_4,k_3,k_2,k_1) =
u(k_4,k_3,k_2,k_1)- 2 g(k_1,k_3) g(k_2,k_4) D(k_1-k_3) \, ,
\label{eq:initial}
\end{eqnarray}
where
\begin{eqnarray}
D(q) = \omega_{{\bf q}}/(\omega^2+\omega_{{\bf q}}^2)
\end{eqnarray}
is the phonon propagator. Here we consider the case of a circular Fermi
surface and anisotropic Einstein
phonons, with $g(k,k^{\prime})=g_0$ and $\omega_{{\bf q}}=\omega_E$. In what 
follows, it is also convenient to define the dimensionless electron-phonon
coupling 
constant  
\begin{eqnarray}
\lambda=2N(0)g_0^2/\omega_E \, . 
\end{eqnarray}

In the Kadanoff-Wilson approach to RG the flow equations are obtained by
computing the corrections to the couplings of the theory as the energy modes
in a energy shell between $\Lambda$ and $\Lambda+d\Lambda$ are integrated
out. The momenta, frequencies, and fields are then rescaled in such a way that
the quadratic terms remain unchanged. This second step presents a difficulty
in the electron-phonon problem. The momenta of the electrons scale
differently in the directions parallel and perpendicular to the Fermi surface,
while the momenta of the phonons scale isotropically. In our RG treatment of
the electron-phonon problem \cite{Tsai}, we use the quantum field theory
version of the RG, with no rescaling. In this approach, corrections to the
vertices of the model are written in term of running, {\it i. e.}
cut-off--dependent, 
coupling functions. The RG flow equations for these running couplings are
then obtained by imposing that the vertices are cut-off independent. 

We start with the two-point vertex, which at one loop is given by \cite{Tsai}
\begin{eqnarray}
\Gamma^{(2)}(\Sigma(\omega, {\mathbf k})) = \Sigma_{\ell}(\omega,{\mathbf k}) -
\int_{-\infty}^{\infty} \frac{d\Omega}{2\pi} \int_{(\Lambda)} 
\frac{d^2{\mathbf k}^{\prime}}{(2\pi)^2} 
\frac{e^{i\Omega 0_+}}{i\Omega - \epsilon_{\mathbf k^{\prime}} -
  \Sigma_{\ell}(\Omega, 
  {\mathbf k}^{\prime})} \tilde{u}(k, k^{\prime}, k, k^{\prime})
\label{eq:gamma2}
\end{eqnarray}
where $k=(\omega,{\mathbf k})$ and $k^{\prime}=(\Omega,{\mathbf
  k}^{\prime})$, and 
$\Sigma_{\ell}(\omega,{\mathbf k})$ is the electronic self-energy which now
flows under the RG, and the RG parameter $\ell = \log(\Lambda_0/\Lambda)$, so
that $d\ell=-d\Lambda/\Lambda$. 
The RG equation for the electron self-energy is obtained by
imposing the condition of renormalizability of the theory, that is, 
\begin{eqnarray}
\frac{d \Gamma^{(2)}}{d\ell} = 0 \, ,
\end{eqnarray} 
where $\ell = \log (\Lambda_0/\Lambda)$ and $d \ell = - d\Lambda/\Lambda$.

Notice that the interaction that appears in 
Eq. (\ref{eq:gamma2}), $\tilde{u}(k,k^{\prime},k,k^{\prime})$,  only describes the
forward scattering channel ($k_4=k_1=k$, $k_3=k_2=k'$) which, in the large
$N$ limit,  
does not flow under RG \cite{Tsai}.  
Therefore, it can be substituted by its unrenormalized value given by
Eq. (\ref{eq:initial}).  
The electron-electron part of the interaction, $u(k,k^{\prime},k,k^{\prime})
= u_0$ contributes only to the real part of the self-energy and gives a shift
in the chemical potential that can be reabsorded into the definition of the
chemical potential \cite{shankar}. The electron-phonon part 
for $\omega > 0$, leads to the following RG equation for the imaginary part of
the self-energy, $\Sigma^{''}(\omega)$:   
\begin{eqnarray}
\frac{\partial}{\partial \ell} \Sigma^{''}_{\ell}( \omega)
= - \int_{-\infty}^{+\infty} \frac{d \Omega}{\pi} 
\lambda \omega_E D(\Omega-\omega)  
\frac{\Lambda_{\ell} (\Omega-\Sigma^{''}_{\ell}(\Omega))}{
(\Omega-\Sigma^{''}_{\ell}(\Omega))^2 + \Lambda_{\ell}^2} \ \ ,
\label{eq:rgself}
\end{eqnarray}
where $\Lambda_{\ell} = \Lambda_0 e^{-\ell}$ (where $\Lambda_0 < E_F$ is the
cut-off in the beginning of the RG flow and it is assumed to be much larger
than the other energy scales in the problem). There is no dependence of
$\Sigma^{''}_{\ell}$ on the direction of the momentum ${\mathbf k}$ because
we are considering the case of isotropic Fermi surface, and the dependence on
the magnitude of ${\mathbf k}$ is irrelevant \cite{shankar}. 
It is convenient to write $\Sigma^{"}(\omega) = [1-Z(\omega)]\omega$, 
and the solution of Eq. (\ref{eq:rgself}) becomes:
\begin{eqnarray}
Z_{\ell}(\omega)  = 1  
 + \frac{1}{\omega} \int_{-\infty}^{+\infty} \frac{d \Omega}{2}  
 \lambda \omega_E D(\Omega-\omega) \Omega
 F_{\ell}(\Omega) 
\label{eq:self}
\end{eqnarray}
 where 
\begin{eqnarray}
F_{\ell}(\omega) = \frac{2}{\pi} \int_{W_{\ell}(\omega)}^{\infty}
\frac{dW^{\prime}}{\left(Z_{W^{\prime}}(\omega)/Z_{W}(\omega)\right)^2
  \omega^2 + W^{\prime 2}} \ \  ,
  \label{fe}
\end{eqnarray}
where we have introduce $W_{\ell}(\omega) = \Lambda_{\ell}/Z_{\ell}(\omega)$
is the new renormalized cut-off running scale. 
The dependence of $Z_W(\omega)$ on $W$ is weak (see below) and therefore we can write
$Z_W(\omega) \approx Z_{W^{\prime}}(\omega)$.  In this case (\ref{fe})
can be solved at once:
\begin{eqnarray}
F_{\ell}(\omega) \approx \frac{1}{|\omega|} \left[1-\frac{2}{\pi} \tan^{-1}
  \left(\frac{W_{\ell}(\omega)}{|\omega|}\right)\right] \, .
\label{bigf}
\end{eqnarray}
In what follows we will be only interested in the low frequency behavior, in which
case $W_{\ell}(\omega)$  can be safely replaced by $W_{\ell}(\omega=0)$, the
expression for $Z_{\ell}(\omega)$ can be easily evaluated from (\ref{eq:self}) to give
\begin{eqnarray}
Z_{\ell}(\omega) = 1 - \lambda \frac{\omega_E}{\omega}
\left[\tan^{-1}\left(\frac{W_{\ell}+\omega_E}{|\omega|}\right) -
  \frac{\pi}{2}\right]  \, .
\label{big_Z}
\end{eqnarray}
Notice that indeed the dependence of $Z_W$ on $W$ is weak, as assumed previously. 
In the static limit, $\omega \to 0$, one obtains 
\begin{eqnarray}
Z_{\ell}(0) = 1 + \lambda \omega_E/(W_{\ell}+\omega_E) \, .
\end{eqnarray}

So far we have considered the renormalization of the self-energy. 
The renormalization of the interaction in the Cooper channel ($k_4=-k_3$ and $k_2=-k_1$) 
can be obtained in a completely analogous way. 
Since we are considering the case of a circular Fermi surface, we can focus
on the $s$-wave component of the BCS interaction
$\tilde{v}(\omega_1,\omega_3) \! = \! N(0) \int d\theta_1/2\pi
\int d\theta_3/2\pi \tilde{u}(-k_3,k_3,-k_1,k_1)$. 
The RG equation for this interaction can be shown to be \cite{Tsai}:
\begin{eqnarray}
\frac{\partial}{\partial \ell} \tilde{v}(\omega_1,\omega_2,\ell)
= - \int_{-\infty}^{+\infty} \frac{d \omega}{\pi}
\frac{\Lambda_{\ell}}{[\omega-\Sigma_{\ell}^{''}(\omega)]^2
+\Lambda_{\ell}^2} \tilde{v}(\omega_1,\omega,\ell) \, 
\tilde{v}(\omega,\omega_2,\ell) \, ,
\label{main_rg}
\end{eqnarray}
where the initial condition for the flow is given by
$\tilde{v}(\omega_1,\omega_3,\ell=0) = u_0 - \lambda
  \omega_E D(\omega_1-\omega_3)$. Equation (\ref{main_rg}) can be
 written in matrix equation as:
\begin{eqnarray}
\frac{\partial {\bf U}}{\partial \ell} = - {\bf U} \cdot {\bf M} \cdot {\bf
  U}
\label{main_rg_matrix}
\end{eqnarray}
where 
\begin{eqnarray}
U_{ij}(\ell) &=& \tilde{v}(\omega_i,\omega_j,\ell)
\\  
M_{ij}(\ell) &=& \Lambda_{\ell}\delta_{ij}/
[(\omega_i-\Sigma^{''}_{\ell}(\omega_i))^2 + \Lambda_{\ell}^2] \, .
\end{eqnarray}
Formally, we can rewrite (\ref{main_rg_matrix}) as
\begin{eqnarray}
\frac{\partial {\bf U}^{-1}}{\partial \ell} = {\bf M}
\label{inverse}
\end{eqnarray} 
since ${\bf U}^{-1} \partial_{\ell} {\bf U} {\bf U}^{-1} = -
\partial_{\ell}{\bf U}^{-1}$. The solution of (\ref{inverse}) reads:
\begin{eqnarray}
{\bf U}^{-1}(\ell) = {\bf U}^{-1}(0) + {\bf P}(\ell)
\end{eqnarray}
 where 
\begin{eqnarray}
{\bf P}(\ell) = \int_0^{\ell} d\ell' {\bf M}(\ell') \, ,
\end{eqnarray}
which can inverted to give:
\begin{eqnarray}
{\bf U}(\ell) =  
[{\bf 1} + {\bf U}(0) \cdot {\bf P}(\ell)]^{-1} \cdot {\bf U}(0) \, .
\label{uell}
\end{eqnarray}

Eq.(\ref{uell}) allows the study of instabilities of the Fermi liquid
state towards superconducting instabilities. The instabilities
occur when one of the couplings diverges under the RG flow.
These instabilities happen at some finite energy (or temperature) 
scale $\ell_c$ at which one of the eigenvalues of the
coupling matrix ${\bf U}(\ell_c) $ diverges at $\ell = \ell_c$.
Notice that the condition for the instability can be written as:
\begin{eqnarray}
\det[{\bf 1}+{\bf U}(0)\cdot{\bf P}(\ell_c)]=0 \, .
\end{eqnarray} 
Hence, the problem reduces to the calculation of the zeros of a determinant
or,  equivalently, 
to the problem of finding the zero eigenvalue of the
matrix $\left[{\bf 1} + {\bf U}(0) \cdot {\bf P}\right]$:
\begin{eqnarray}
[{\bf 1} + {\bf U}(0) \cdot {\bf P}(\ell_c)] \cdot {\bf f} =0
\label{eigenvalue_problem}
\end{eqnarray}
where the $f_i$ is the eigenvector of the problem. 
Eq. (\ref{eigenvalue_problem}) can be written explicitly as:
\begin{eqnarray}
f(\omega) = - \int_{-\infty}^{+\infty} \frac{d \Omega}{\pi}
\int_{\Lambda_c}^{\Lambda_0} 
d\Lambda \frac{1}{[\Omega Z_{\ell}(\Omega)]^2+\Lambda^2}
 \left[u_0 - \frac{\lambda \omega_E^2}{(\omega-\Omega)^2+\omega_E^2}\right]
f(\Omega) \, .
\label{int_eq_2}
\end{eqnarray}
Here we can, similarly to the expression for the self-energy (\ref{eq:self}),
write 
\begin{eqnarray}
Z_{\ell_c}(\omega) \phi_{\ell_c}(\omega) = 
- \int_{-\infty}^{+\infty} \frac{d \omega'}{2}
\left[u_0-\frac{\lambda \omega_E^2}{(\omega-\omega')^2+\omega_E^2}\right]
F_{\ell_c}(\omega') \, \, 
\phi_{\ell_c}(\omega') \, ,
\label{appr_1}
\end{eqnarray}
where we have  defined $\phi_{\ell_c}(\omega) = f_{\ell_c}(\omega)/Z_{\ell_c}(\omega)$. 
Equations (\ref{big_Z}) and (\ref{appr_1}) determine the energy scale $W_c = W_0 e^{-\ell_c}$
at which the renormalization group equations for the scattering in the Cooper channel
diverge as one renormalizes the problem from high to low energies. Below this energy scale
the Fermi liquid breaks down and superconductivity sets in. Thus, we can associate 
$W_c$ with the superconducting gap, $\Delta$. In fact, we will show that the equation for
$W_c$ gives exactly the same result obtained from Eliashberg's theory for strongly coupled
superconductors \cite{eliashberg}. 

\section{Solution of the RG flow equations}
\label{sec:flows}

Traditionally the superconducting temperature has been calculated directly from
Eliashberg's equations \cite{eliashberg}. Nevertheless, the
formalism presented in the previous section allows the solution of the
problem by 
solving equations (\ref{big_Z}) and (\ref{uell}), instead. The advantage of
such 
a procedure is clear since it is not necessary to solve integral equations. 

It is interesting to investigate how the coupling matrix evolves under the RG
flow in different regimes since it provides an insight on how to solve the
problem analytically in some asymptotic limits. The simplest case occurs
when there are no phonons present  (as discussed in Ref.\onlinecite{shankar}) 
where a full analytical solution is possible. 
For the case of $\lambda =0$, Eq. (\ref{big_Z}) gives $Z_{\ell}=1$ and (\ref{uell})
becomes independent of the external frequency and therefore we must have 
$\phi(\omega) = C$ is a non-zero constant. In this case (\ref{uell}) gives:
\begin{eqnarray}
1 &=& - u_0  \int_{-\infty}^{+\infty} \frac{d \omega'}{2} F_{\ell_c}(\omega') 
\nonumber
\\
 &=& -u_0 \ln(W_0/W_c)
\end{eqnarray}
where we have used (\ref{bigf}). Notice that because $W_c < W_0$ the above
equation only has solution if $u_0<0$, that is, if the unrenormalized electron-electron
interactions are attractive to start with. The solution of the above equation
gives:
\begin{eqnarray}
\Delta = \Lambda_0 e^{1/u_0} \, ,
\end{eqnarray} 
where $W_0 = \Lambda_0$ since $Z=1$. 
Since, in our case we assume $u_0>0$, no superconducting instability
can exist in the absence of phonons.

\begin{figure}[htb]
\includegraphics[scale=0.9]{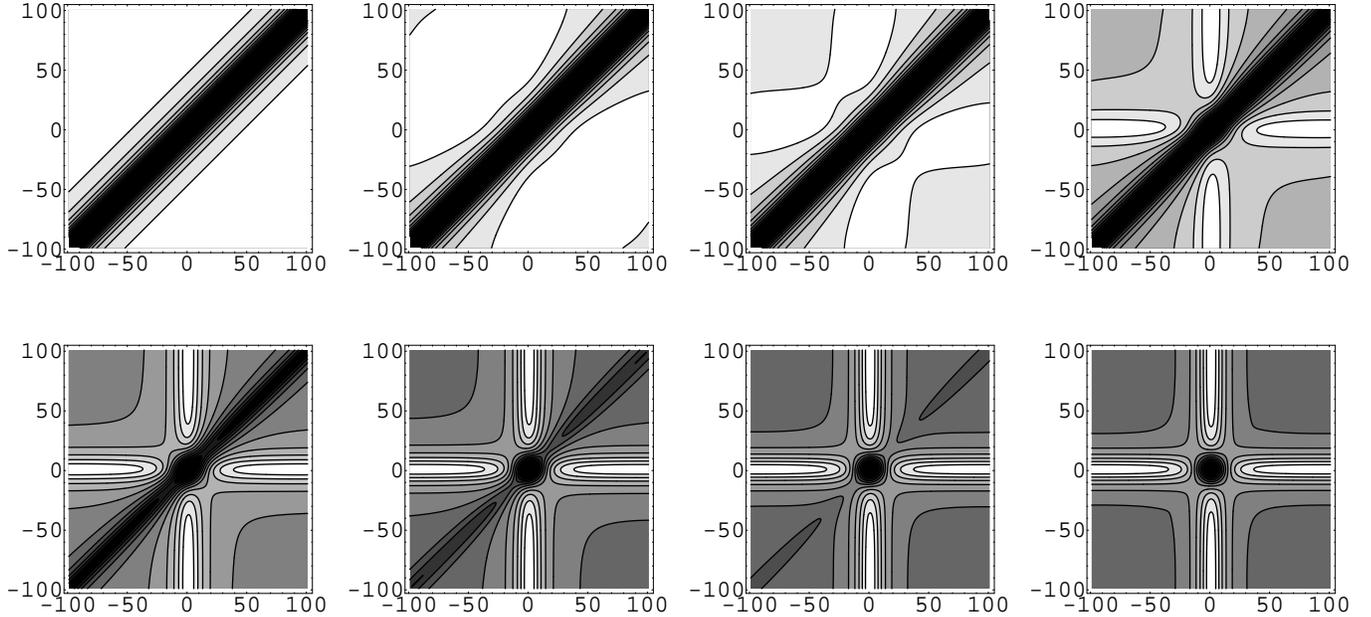}
\caption{Plots of the $N$x$N$ matrix $U$ at different RG scales
  $\ell$. Here the number of frequency divisions $N\!=\!200$, and the value
  of the parameters used are $\lambda\!=\!0.3$,
  $\Lambda_0\!=\! 100$, $\omega_E\!=\! 10$, 
  $u_0\!=\! 0.1$. Panels correspond to $\ell$ = 0, 2.5, 3, 5, 6.5, 6.9, 7.1,
  and 7.19.} 
\label{fig:weak}
\end{figure}

Instead of only focusing on the solution of the RG equation at the
instability, at which point the zero eigenvalue condition
(\ref{eigenvalue_problem}) holds, we have also obtained the full solution for
the RG flow by solving (\ref{eq:self}) and (\ref{uell}) numerically. Fig.
\ref{fig:weak} shows the evolution of ${\bf U}(\ell)$ with $\ell$. Each panel
in Fig. \ref{fig:weak} represents the $N$x$N$ matrix ${\bf U}(\ell)$ at a given
$\ell$. The frequency has been cut-off so that $|\omega| < \Lambda_0$ and has
been discretized in $N=200$ divisions in this interval. In a typical flow in
the weak-coupling limit (in the case of Fig. \ref{fig:weak} $\lambda = 0.3$,
$\omega_E=10$, and $u_0=0.1$), ${\bf U}$ flows from the initial condition where
$\tilde{v}(\omega,-\omega)$ are the largest couplings to the instability point
where the couplings that first diverge are the $\tilde{v}(\omega_i,\omega_j)$
where $|\omega_i|, |\omega_j| < \omega_E$. 

We can clearly see from Fig.\ref{fig:weak} that, as the RG develops,
the matrix ${\bf U}$, which at the beginning of the RG has its largest
elements along  
the diagonal, acquires a cross-like shape, at which point its largest matrix
elements occur near the origin where the frequencies are very small. The
region with largest matrix elements occurs within the dark circle of size
$\omega_E$ around the origin. Moreover, within this circle the matrix
elements are essentially constant showing that there is very little frequency
dependence if the frequencies are smaller than $\omega_E$. 
Hence, $\omega_E$ is the scale that separates high from low
energy physics in the weak coupling regime. Thus, the full frequency dependence
of the phonon propagator is irrelevant and we can safely approximate:
\begin{eqnarray}
\lambda \omega_E D(\omega) \approx \left\{
\begin{array}{ll}
\lambda & \hskip 0.5cm {\rm if} \ \ \ |\omega| < \omega_E\\
0 & \hskip 0.5cm {\rm if} \ \ \ |\omega| > \omega_E\\
\end{array}
\right. \, .
\label{phonon_prop_app}
\end{eqnarray}
Hence, from Eq. (\ref{big_Z}) we have 
$Z_{\ell}(\omega) \approx 1+ \lambda$ which is independent of $\ell$.

Even tough the most divergent elements of the matrix ${\bf U}$ are the ones
at small frequencies, it is important to include all the matrix elements
since the RG equation couples low and high frequency processes.
Again, making use of the fact that $\omega_E$ is the energy scale that
separates high and low energy physics in this limit, 
we can propose the  {\it ansatz}: 
\begin{eqnarray}
\phi(\omega) = \left\{
\begin{array}{ll}
\phi_0 & \hskip 0.5cm {\rm if} \ \ \ |\omega| < \omega_E\\ 
\phi_{\infty} & \hskip 0.5cm {\rm if} \ \ \ |\omega| > \omega_E\\ 
\end{array}
\right.
\label{ansatz_weak}
\end{eqnarray}
where $\phi_0$ and $\phi_{\infty}$ are unknowns that to be calculated. 
Substituting (\ref{phonon_prop_app}), (\ref{ansatz_weak}), and the expression
for $Z_{\ell_c}$ into (\ref{uell}) one finds:
\begin{eqnarray}
\left[1+\lambda+(u_0 - \lambda) \, h(\omega_E/\Lambda_c)\right] \phi_0 +
(1+\lambda) (\ell_c - h(\omega_E/\Lambda_c)) \phi_{\infty}&=&0 \, ,
\nonumber
\\
u_0 h(\omega_E/\Lambda_c) \phi_0 + (1+\lambda) \left[1+(\ell_c
  -h(\omega_E/\Lambda_c))u_0\right] \phi_{\infty} &=& 0  \, ,
\label{eq_weak}
\end{eqnarray}
where
\begin{eqnarray}
h(z) = \frac{2}{\pi} \int_0^{z} dx \, \frac{\tan^{-1}(x)}{x} \, ,
= \frac{z}{2 \pi} \Phi(-z^2)
\label{fc}
\end{eqnarray}
and where $\Phi(x)$ is a LerchPhi function, $\Phi(x) = \sum_{k=0}^{\infty}
x^k/(k+1/2)^2$. The solution of (\ref{eq_weak}) is given by the determinant
equation: 
\begin{eqnarray}  
\left|
\begin{array}{ll}
1+\lambda+(u_0 - \lambda) \, h(\omega_E/\Lambda_c) & \hskip 0.5cm
(1+\lambda) (\ell_c - h(\omega_E/\Lambda_c)) \, u_0  \\ 
u_0 \, h(\omega_E/\Lambda_c) & \hskip 0.5cm (1+\lambda)(1+(\ell_c
-h(\omega_E/\Lambda_c))) \, u_0 \\ 
\end{array}
\right| = 0 \, ,
\end{eqnarray}
or equivalently,
\begin{eqnarray}
\lambda u_0 h^2(\omega_E/\Lambda_c) - \lambda(1+u_0+ u_0 \ell_c)
h(\omega_E/\Lambda_c) + (1+\lambda) \, (1 + u_0 \ell_c) = 0 \, ,
\end{eqnarray}
which is a transcendental equation for $\Lambda_c$. The last equation can be
rewritten in a more interesting form by defining $\Lambda_c = x_c \omega_E$
where $x_c$ is the variable of the problem (notice that $\ell_c =
\ln(\Lambda_0/\Lambda_c) = \ell_E + \ln(1/x_c)$ where $\ell_E =
\ln(\Lambda_0/\omega_E)$) in order to get:
\begin{eqnarray} 
h^2(1/x_c) - \left[1/\mu^* +1+\ln(1/x_c)\right] h(1/x_c) + 
\frac{1+\lambda}{\lambda} \left[1/\mu^* + \ln(1/x_c)\right] = 0 \, ,
\label{trans_weak}
\end{eqnarray}
where
\begin{eqnarray}
\mu^* = \frac{u_0}{1+ u_0 \ln(\Lambda_0/\omega_E)} \, ,
\end{eqnarray}
is the renormalized electron-electron interaction at the scale of phonon
energy $\omega_E$. This is the Anderson-Morel potential\cite{Morel}, which
emerges naturally from the solution to the RG flow. 

Note that the
renormalized interaction is really weakly dependent on the bare 
electron-electron interaction since as $u_0$ increases it saturates very fast
at $\mu^* \approx 1/\ln(\Lambda_0/\omega_E)$ which is interaction
independent. In fact, in ordinary superconductors $\mu^*$ varies little from
material to material ($0.1<\mu^*<0.25$) and this can be understood by the
logarithmic dependence with the cut-off $\Lambda_0$ and the phonon frequency
$\omega_E$ (if we estimate $\Lambda_0 \approx 10^4$K while 
$\omega_E \approx 10^2$K we get $\mu^* \approx 0.217$).

When $x_c \ll 1$ ($\Lambda_c \ll \omega_E$) we can approximate $h(1/x_c)
\approx \ln(1/x_c)$ and Eq. (\ref{trans_weak}) becomes
\begin{eqnarray}
\left(\frac{1}{\mu^*}+1-\frac{1+\lambda}{\lambda}\right) \ln(1/x_c) \approx
\frac{1+\lambda}{\lambda \mu^*}  \, ,
\end{eqnarray}
which leads to
\begin{eqnarray}
W_c \approx \omega_E \, \exp [- (1+\lambda)/(\lambda - \mu^*)]
\end{eqnarray}
which is the MacMillan expression for the superconducting gap at zero
temperature \cite{mc,allen}. 

\begin{figure}[htb]
\includegraphics[scale=0.9]{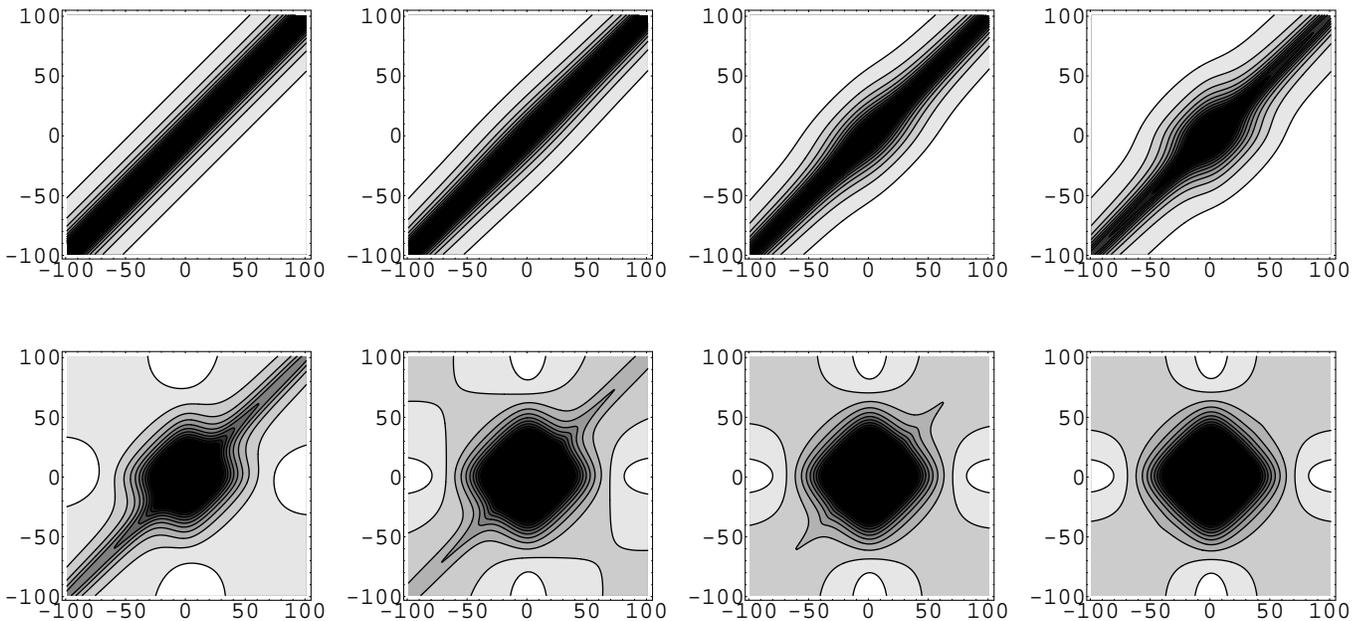}
\caption{Plots of the $N$x$N$ matrix $U$ at different RG scales
  $\ell$. Here the number of frequency divisions $N\!=\!200$, and the value of
  the other parameters are $\lambda\!=\!4$, $\Lambda_0\!=\!100$,
  $\omega_E\!=\! 10$, and 
  $u_0\!=\!0.1$. Panels correspond to $\ell$ = 0, 1, 2, 2.5, 3, 3.13, 3.157,
  3.172. The scale $2W_c \approx 40$ distinguishes the high and low
  frequencies close to $\ell_c$.}
\label{fig:strong}
\end{figure}

In the strong-coupling regime ($\lambda > 1$) the situation is rather different.
Fig. \ref{fig:strong} shows the numerical results for the full solution of
the RG flow for parameters in strong coupling. The form of the
matrix ${\bf U}(\ell)$ is very different from the weak-coupling limit and the
important scale separating the high and low energy physics is $2W_c$.
We have seen that for $\lambda <1$ the characteristic energy scale
that separates high from low energy physics is $\omega_E$. 
When $\lambda > 1$ we expect $W_c > \omega_E$ in which case
$\omega_E$ is not the characteristic energy scale of the instability.
Instead we expect $W_c$ to act as the characteristic energy
scale in separating the low and high energy physics. Thus, it is natural to
make the following {\it ansatz}:
\begin{eqnarray}
\phi(\omega) = \left\{
\begin{array}{ll}
\phi_0 & \hskip 0.5cm {\rm if} \ \ \ |\omega| < 2 W_c\\ 
0 & \hskip 0.5cm {\rm if} \ \ \ |\omega| > 2 W_c\\ 
\end{array}
\right. \, .
\label{ansatz_strong}
\end{eqnarray}
where $\phi_0$ is an unknown. 
Since, as discussed previously, $\mu^*$ is bounded from above,
the electron-electron interaction will play a minimal role in the problem and
can be disregarded.
 
In this regime (\ref{appr_1}) becomes ($\phi_0 \neq 0$):
\begin{eqnarray}
Z_c(0) = 2 \lambda \omega^2_E \int_0^{2 W_c} \frac{d \omega}{\pi}
\int_{W_c}^{+\infty} d\Lambda \frac{1}{(\omega^2+\Lambda^2) \, 
(\omega^2+\omega_E^2)}
\approx \frac{2 \lambda}{\pi} \left( \frac{\pi \omega_E}{2 W_c}
-\frac{\pi \omega_E^2}{4 W_c^2}\right)  \, .
\label{eq_strong_1}
\end{eqnarray}
But from (\ref{big_Z}) we have $Z_{c}(0) \approx 1 +
\lambda (\omega_E/W_c) - \lambda (\omega_E/W_c)^2$, 
which substituted in (\ref{eq_strong_1}) gives
\begin{eqnarray}
W_c \approx \sqrt{\lambda} \omega_E \, ,
\end{eqnarray}
the strong coupling result of Eliashberg 
equation found by Allen and Dynes \cite{allen}.

\section{Discussion and conclusion}
\label{sec:conclusion}

\begin{figure}[htb]
\includegraphics[scale=0.45,angle=0]{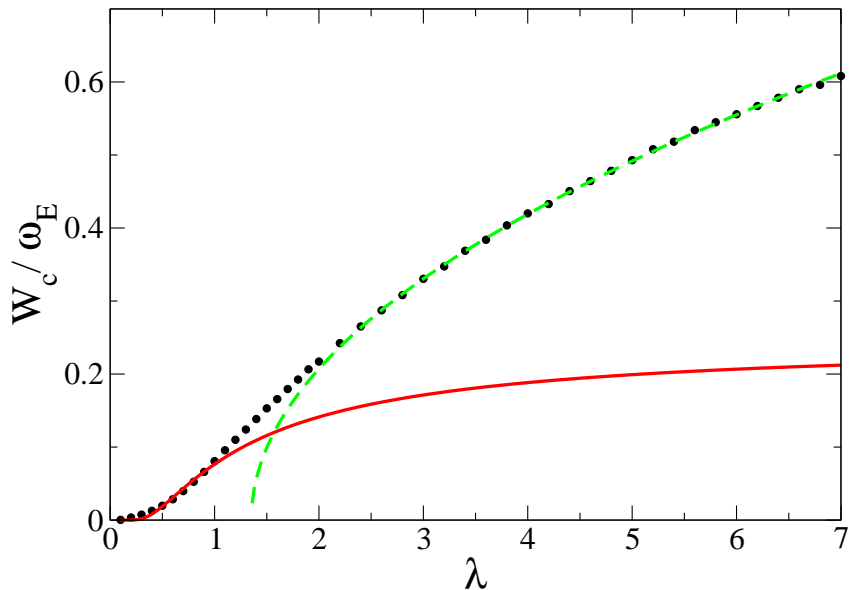}
\caption{Plot of $W_c$ ($\sim $ superconducting gap) in units of
  $\omega_E$ versus $\lambda$ (black circles) ($N=200$,
  $\Lambda_0 = 100$, $\omega_E=10$, $u_0=0.1$). The dashed line is the
fit at small $\lambda$'s using the MacMillan formula and the
solid line is the fit to $\sqrt{\lambda}$ at large $\lambda$'s (Allen-Dynes). }
\label{fig:lamc}
\end{figure}

We have numerically solved the RG equations for the flow of the BCS coupling
matrix for interacting electrons coupled to phonons, for the case of circular
Fermi surface and Einstein phonons. Fig. \ref{fig:lamc}
shows the result for the energy scale of divergence in the Cooper channel,
for a range of $\lambda$, and the expected analytical result at weak and
strong coupling limits. One can clearly see that the approximate analytical
solutions, based on the evolution of the coupling matrix under the RG, give a
very good description of the superconducting gap in the asymptotic
limits. Thus, our RG scheme can used as an alternative way to solve for the
superconducting properties of a Fermi liquid and is in full agreement with
Eliashberg's theory.

Unlike mean-field approaches like Eliashberg theory, the RG approach, when
solved 
numerically, can be easily extended to any shape of the Fermi surface or
anisotropic phonons. It also addresses the question of competition between
different types of instabilities, as was done extensively for the case
without phonons\cite{zanchi,halboth,marston}. Furthermore, the approach
presented here at zero temperature can be easily extended to finite
temperature.

\acknowledgments
We thank J.~Carbotte, A.~Chubukov, C.~Chamon, J.~B.~Marston,
G.~Murthy, and M.~Silva-Neto for illuminating discussions
and the Aspen Center for Physics for its hospitality during the
early stages of this work. A.~H.~C.~N. was supported
by NSF grant DMR-0343790. R.~S. was supported by NSF grant
DMR-0103639.

\end{document}